\documentclass[12pt]{iopart}
\usepackage{citesort}
\usepackage{iopams} 
\usepackage{graphicx}

\begin{document}

\title{Field-dependent AC susceptibility of itinerant ferromagnets}

\author{M D Vannette}
\author{R Prozorov}

\address{Ames Laboratory and Iowa State University, Dept. of Physics \& Astronomy, Ames, IA 50011}

\ead{vannette@iastate.edu}
\ead{prozorov@ameslab.gov}

\begin{abstract}

Whereas dc measurements of magnetic susceptibility, $\chi$, fail to distinguish between local and weak itinerant ferromagnets, radio-frequency (rf) measurements of $\chi$ in the ferromagnetic state show dramatic differences between the two.  We present sensitive tunnel-diode resonator measurements of $\chi$ in the weak itinerant ferromagnet ZrZn$_2$ at a frequency of 23 MHz.  Below Curie temperature, $T_C \approx 26$ K, the susceptibility is seen to increase and pass through a broad maximum at approximately 15 K in zero applied dc magnetic field. Application of a magnetic field reduces the amplitude of the maximum and shifts it to lower temperatures. The existence and evolution this maximum with applied field is not predicted by either the Stoner or self-consistent renormalized (SCR) spin fluctuations theories. For temperatures below $T_C$ both theories derive a zero-field limit expression for $\chi$. We propose a semi-phenomenological model that considers the effect of the internal field from the polarized fraction of the conduction band on the remaining, unpolarized conduction band electrons. The developed model accurate describes the experimental data.  

\end{abstract}

\submitto{\JPCM}
\maketitle

\begin{figure}[htbp]
\begin{center}
\includegraphics[width=9cm]{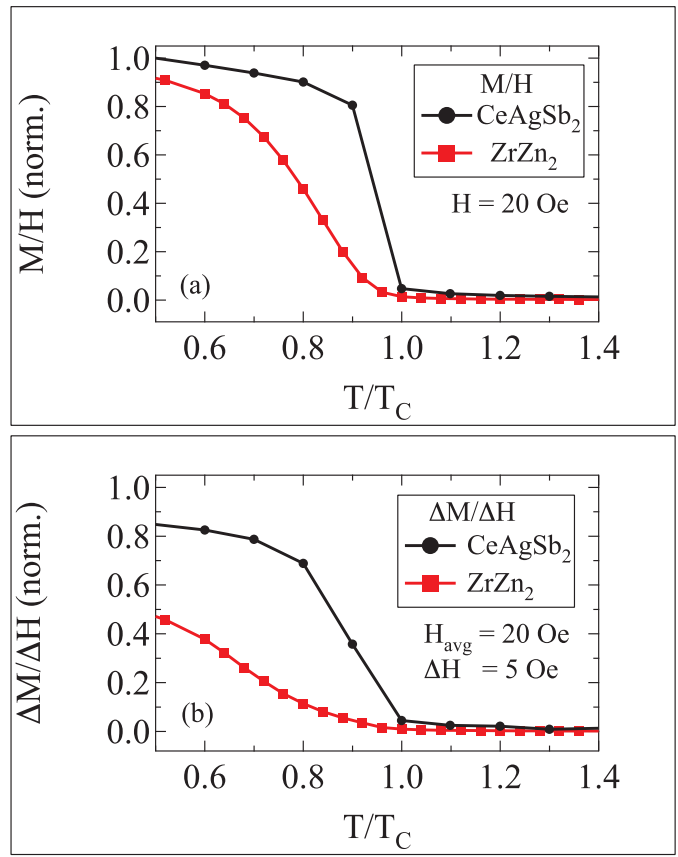}
\end{center}
\caption{\label{dcmom}(Color online) Comparison of normalized measurements of dc susceptibility of the 4$f$ local moment ferromagnetic CeAgSb$_2$ (black circles, $T_{C}\approx9.8$ K)  and the weak itinerant ferromagnet ZrZn$_2$ (red squares, $T_{C}\approx27$ K) on a reduced temperature scale.  Panel (a) is the conventional dc susceptibility, $M/H$.  Panel (b) is the result of dc delta measurements as explained in the text.  Both sets of measurements were carried out with $H_{avg}=20$ Oe.  Normalization is relative to the 5 K value of $M/H$ and performed for each sample separately.}
\end{figure}

DC measurements of magnetic susceptibility fail to distinguish between local and itinerant ferromagnets.  A common method of determining $\chi_{dc}$ is measuring the magnetic moment and then dividing by the applied field, $H$.  However, this is only applicable provided the magnetization is linear in $H$ from $H=0$ up to the measurement field.  For magnetically soft or small moment ferromagnets, this criterion may not be satisfied. Perhaps a more careful method is to measure $M$ in two slightly different magnetic fields and then calculate $\Delta M/\Delta H$ as shown in Fig.~\ref{dcmom}.  Further, due to limited sensitivity, DC measurements are usually conducted in  significant magnetic fields, on the order of 1-10 Oe. In exceptionally soft materials these fields may be sufficient to smear certain zero-field features.  

In itinerant systems the situation is even more complicated. In order to deduce the size of a magnetic moment per ion, one has to be in a saturation regime by applying a large field. However, this tells nothing about the magnitude of this moment in zero field that, in itinerant systems, is field - dependent. Yet, even with a conventional definition, observation of a magnetic moment that is a fraction of a Bohr magneton per ion is not bulletproof evidence for the itinerant nature of magnetism. For example, the local-moment metallic rare-earth compound CeAgSb$_2$ \cite{myers-1999} and the insulating titinate YTiO$_3$ \cite{garrett-1981} both possess a fractional magnetic moment per ion. Possible explanations of such fractional local moments could be a canted antiferromagnetic structure as may occur in YTiO$_{3}$ \cite{ulrich-2002}, or crystalline electric field effects as has been proposed for CeAgSb$_{2}$ \cite{araki-2003}.

Unlike dc measurements, ac susceptibility can be measured in much lower magnetic fields. It became a valuable technique in studying magnetic materials. However, interpretation of the results, especially in itinerant systems is complicated.
For example, low frequency ac susceptibility measurements on the insulating two dimensional ferromagnet K$_{2}$CuF$_{4}$ \cite{suzuki-1981} is strikingly similar to that measured in palladium slightly doped with manganese \cite{ho-1981} as well as in an Fe-Ni-B-Si alloy \cite{drobac-1996}.  While it is clear that the insulating compound is a local moment, the nature of the latter two is open to debate.  It is suggested in these works that the similarities in the data is due to a combination of demagnetization effects and domain wall motion. 

In this contribution we report radio frequency temperature dependent ac susceptibility of the well-studied commonly accepted itinerant ferromagnet ZrZn$_2$, and examine its evolution with an applied magnetic field. We then present a semi-phenomenological model that describes our data.
Low frequency ac susceptibility measurements on ZrZn$_2$ have been reported \cite{yelland-2005-2}, however, as the focus of that work was not the ferromagnetism of the compound, the only presented data is for $T<2$ K.

\begin{figure}[htbp]
\begin{center}
\includegraphics[width=9cm]{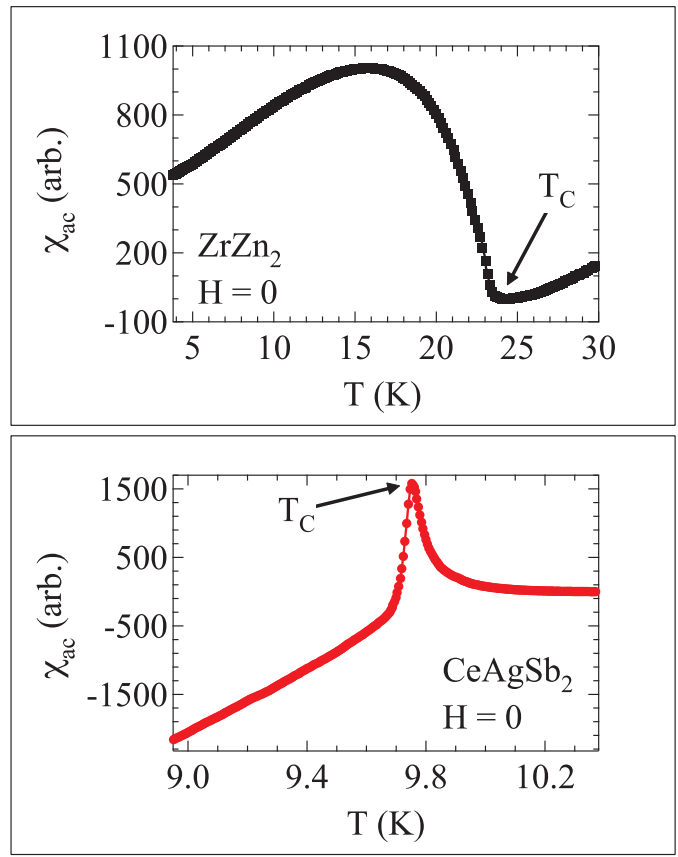}
\end{center}
\caption{\label{comparison}(Color online) Radio frequency susceptibility of the weak itinerant ferromagnet ZrZn$_2$ (top)  and  the $4f$ local moment ferromagnet CeAgSb$_{2}$ (bottom) in zero applied field.   $T_C$ marks the feature at the Curie temperature for each material. The general decrease in the measured $\chi$ (above $T_{C}$ for ZrZn$_{2}$, and below for CeAgSb$_{2}$) is caused by a decrease in resistivity.  Note the different temperature scales in each plot.}
\end{figure}

Recently, it has been shown \cite{vannette-2008} \cite{vannette-2008-2} that rf measurements of ac magnetic susceptibility, $\chi_{ac}$, seem to distinguish between local moment and itinerant ferromagnetism.  Further, in Ref.~\cite{vannette-2008} an explanation ruling out the demagnetzation or magnetic domains effects was presented. It is clear from Fig.~\ref{comparison} that the rf susceptibility of the weak itinerant ferromagnet ZrZn$_2$ \cite{knapp-1971} is distinctly different from that of the $4f$ local moment CeAgSb$_2$ \cite{myers-1999}.  The purpose of this work is to present an effective Weiss-type model to describe the  data derived from itinerant ferromagnets.

The design and operation of a tunnel-diode resonator (TDR) are described in detail elsewhere \cite{vandegrift-1975} \cite{srikanth-1999} \cite{prozorov-2000}.  The device is built around a tunnel diode, a semiconducting device with a voltage bias region of negative differential resistance.  Biasing to this voltage region allows the tunnel diode to drive an $LC$ tank circuit at its natural resonant frequency. In magnetic measurements, a sample is placed in the coil of the tank circuit thereby changing the total inductance, and, hence, the tank circuit's resonant frequency.  It can be shown \cite{clover-1970} that the frequency shift of the tank circuit is directly proportional to the magnetic susceptibility, $\chi$, of the sample in the coil as
\begin{equation}
{\frac{\Delta f }{{f_0}}} \approx -{\frac{1}{2}} {\frac{V_s}{{V_c}}}4 \pi
\chi_m.
\end{equation}
Here $V_s$ and $V_c$ are the volumes of the sample and coil, respectively, and $\chi_m$ is the measured susceptibility of the sample.  Careful design and construction allows one to resolve changes in resonant frequency induced by the sample on the order of 1-10 mHz.   The resulting tuned circuit, operating at 10-20 MHz, gives frequency sensitivity on the order of a few parts per billion.  This translates to a typical sensitivity of $10^{-7}-10^{-8}$ change in $\chi$ induced by temperature or magnetic field.  Due to the operating frequency the measured susceptibility is composed of two parts.  The first is due to the magnetic moments in the sample, and may be either para- or diamagnetic depending on the material studied.  The second is due to the screening of an rf field via the normal skin effect in metals.  This screening is a diamagnetic contribution and is a measure of changes in resistivity \cite{prozorov-2006}.  

Radio frequency susceptibility data presented herein were collected in a TDR operating at 23 MHz mounted in a $^{4}$He cryostat.  The design is similar to that presented in Ref.~\cite{vannette-2008}.  The temperature of the sample can be varied from 3 to 100 K and a dc magnetic field up to 2.5 T coaxial with the rf excitation field ($\sim 20$ mOe) can be applied with a superconducting magnet.  The magnet is mounted inside the vacuum can of the cryostat resulting in no trapped magnetic field at the beginning of each run.  As the effects studied herein are completely suppressed by fields on the order of 500 Oe, any such trapped magnetic field could affect the data.  Single crystal samples were used in this study.  The CeAgSb$_{2}$ sample was prepared as described in Ref.~\cite{myers-1999}, while the ZrZn$_{2}$ sample was prepared as described in Ref.~\cite{dereotier-2006}.

Figure \ref{dcmom} compares temperature dependent dc susceptibility for the local moment CeAgSb$_{2}$ with that for the itinerant ZrZn$_{2}$ as measured in a \textit{Quantum Design} MPMS-5.  Two different techniques were used to determine these susceptibilities.  Panel (a) shows the usual $\chi_{dc}$ where the magnetic moment is measured in a small applied field ($H=20$ Oe) and $M/H$ is calculated.  While this method is appropriate for temperatures well above $T_{C}$ where magnetization is linearly dependent on field over a fairly large field range, it should be expected to fail in the ferromagnetic state because $M$ is not necessarily linear in $H$ all the way down to $H=0$.  Bearing this in mind, a delta measurement of $\chi_{dc}$ was performed (results in panel (b)) as follows.  Magnetic moment versus temperature was measured first in a 17 Oe field and then in a 22 Oe field.  The difference in the resulting moment was divided by the 5 Oe difference in applied fields to determine $\Delta M/\Delta H$.  The advantage of this method is that it only requires approximate linearity in $M(H)$ over the 5 Oe window defined by the upper and lower fields.  Thus, it can be expected to approximate $\chi=\frac{dM}{dH}$ more closely.  Obviously, a smaller $H$ window is more likely to conform to the linear $M(H)$ approximation.

While the delta measurement results in a lower susceptibility in both samples, both dc techniques produce quite similar $\chi(T)$ curves.  This is to be contrasted with the results of zero field radio frequency susceptibility versus temperature as shown in Fig.~\ref{comparison}.  Whereas the local moment system shows a sharp, well defined peak in $\chi_{ac}$ at the Curie temperature, the itinerant system exhibits a broad maximum well below $T_{C}$.

Conventional theories of itinerant ferromagnetism fail to predict the behavior seen in the TDR data of ZrZn$_2$.  The development of the Stoner theory \cite{stoner-1938} was driven by a desire to understand how a fractional Bohr magneton magnetic moment could be created in nickel.  The failures of Stoner theory, \textit{i.e.} Curie temperatures that are too high and the lack of a Curie-Weiss susceptibility above $T_{C}$, were impetus for the development of the spin fluctuation theory of Moriya and Kawabata \cite{moriya-1973}.  While spin fluctuation theory does indeed predict a Curie-Weiss type paramagnetic state and largely correct the Curie temperatures, neither it nor the Stoner theory adequately describe the broad maximum seen in the rf data.  Indeed, both theories derive a strictly zero field limit of $\chi$ just below $T_C$ of the same form
\begin{equation}
\label{stonersf}
\chi(T<T_C)= \chi_{0}\left(1-\left( \frac{T}{T_C}\right) ^{n}\right)^{-1}.
\end{equation}
The difference between the two theories is the value of $n$.  For Stoner theory $n=2$ while for spin fluctuations $n=1$ \cite{mohn-2003}.  The Stoner theory does predict a nonzero $\chi$ at $T=0$ and $H=0$ \cite{wohlfarth-1968}, however there is no prescription for how the susceptibility should evolve from $T_{C}$ to 0 K.

A zero field limit calculation, however, is not representative of a ferromagnetic system below $T_C$.  As the system begins to order there is a non-zero field in the sample from the ordered moments themselves.  In itinerant systems this mean field should continue to increase as $T$ decreases and the fraction of spin polarized conduction electrons increases.  The increase in the itinerant mean field may be expected to be more dramatic than in a local moment mean field because in the former there are physically more magnetic carriers as temperature decreases, while in the latter there is merely less thermal randomization of the moment directions.  To account for the self-field we propose a modified Brillouin function for a spin-1/2 system in a magnetic field.  We choose a spin-1/2 system because in itinerant systems it is the single electron spin that is of interest.

\begin{equation}
\label{moment}
m^{*}(t,h)=m^{*}_{0} \tanh\frac{h}{1-t^n}.
\end{equation}

In the above equation, $t=\frac{T}{T^{*}}$ where $T^{*}$ is a characteristic temperature not necessarily equal to $T_C$ and $h$ is a dimensionless field term.  It should be noted that this form does not represent the magnetization of the sample.  Rather, $m^{*}$ may be taken as some measure of the unpolarized fraction of the conduction band.  Differentiating with respect to $h$ gives

\begin{equation}
\label{fiteq}
\chi(t,h)=\frac{\chi_0}{1-t^n}\cosh^{-2}\frac{h}{1-t^n}.
\end{equation}
In the limit $h\rightarrow 0$, this reduces to Eq.\ \ref{stonersf} if $T^{*}\rightarrow T_{C}$, despite the fact that we are not beginning with the proper form for the magnetization.  

To account for the resistivity contribution to the measured susceptibility, data collected in a dc field of 1 kOe were subtracted from lower field runs.  This field was sufficiently large to completely suppress the maximum in $\chi_{ac}$ while being small enough that it is not expected to result in a significant magnetoresistance. 

\begin{figure}[htbp]
\begin{center}
\includegraphics[width=9cm]{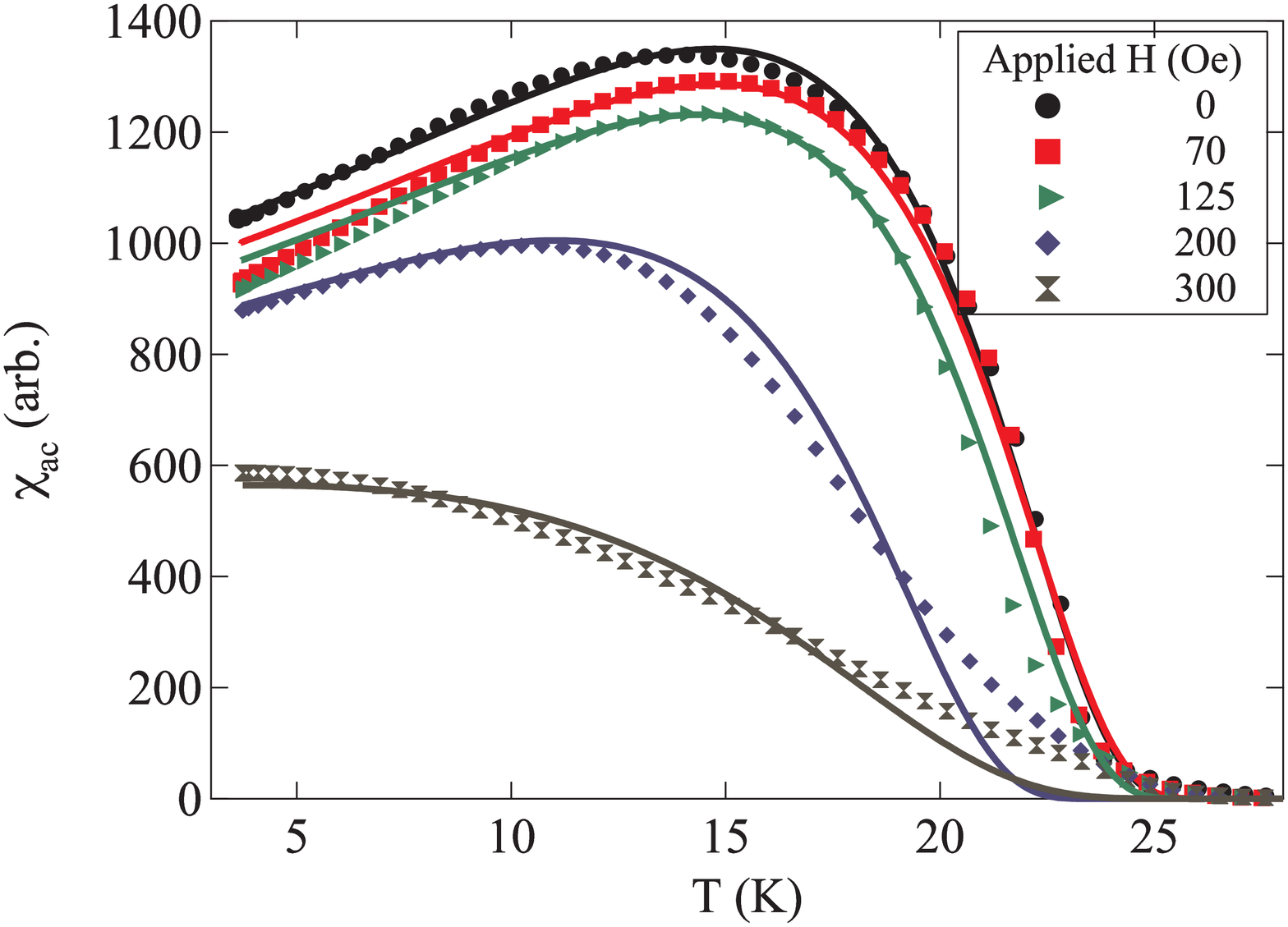}
\end{center}
\caption{\label{fits}(Color online) Comparison of data (points) and fits (solid lines) from the model presented in Eq. \ref{fiteq} with $n=1$.  For clarity only every seventh data point is shown.  In all fits, $R^2$ values were greater than 0.98.}
\end{figure}

\begin{figure}[htbp]
\begin{center}
\includegraphics[width=9cm]{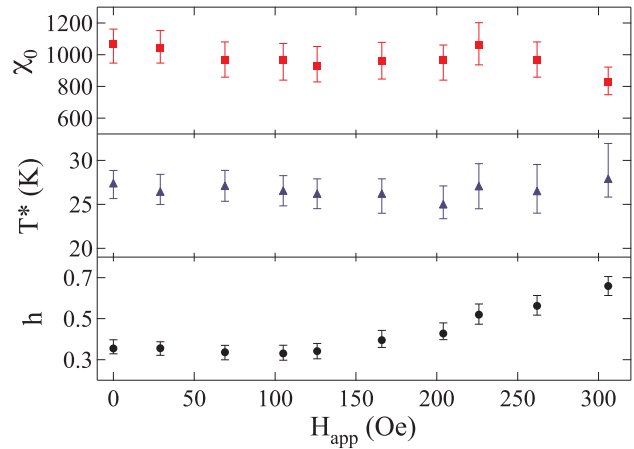}
\end{center}
\caption{\label{fitparams}(Color online) Values of fitting parameters $\chi_{0}$, $T^*$, and $h$ (top to bottom) for ZrZn$_2$ single crystal TDR data as functions of applied field ($H_{app}$) derived from Eq. \ref{fiteq}.  Errors were determined by individually varying the fit parameters until the $R^2$ value dropped below 0.95.}
\end{figure}

Data fits to the model were attempted for various values of $n$.  It was found that $n=1$, corresponding to the spin fluctuations theory, gave the best agreement.  Figure \ref{fits} shows best fits of Eq.\ \ref{fiteq} to the TDR data for ZrZn$_{2}$.  The resulting values of the fitting parameters are shown in Fig.\ \ref{fitparams}.  $\chi_{0}$ decreases with applied $H$.  $T^{*}$ is constant within the errors and it is approximately equal to $T_{C}$.  The value of $h$ is approximately constant up to applied fields of about 125 Oe and thereafter grows monotonically as $H$ is increased.  

In weak itinerant ferromagnets, like ZrZn$_{2}$, the susceptibility in the ferromagnetic state is dominated by contributions from the polarized fraction of conduction electrons and the unpolarized fraction.  This second contribution comes from a band with a large Stoner enhancement, so it should be expected to have a correspondingly large susceptibility.  We suggest that our model accounts for the behavior of the unpolarized, fluctuating portion of the conduction band.

The operating frequency in this study is commensurate with domain-wall resonance techniques \cite{saitoh-2004}.  However, in ZrZn$_{2}$ dc magnetization measurements suggest that single crystals are forced into a single domain in fields as small as 30 Oe \cite{yelland-2005} effectively ruling out any domain wall motion.

In conclusion, we have presented rf susceptibility measurements on the weak itinerant ferromagnet ZrZn$_{2}$ in various applied dc fields.   The zero field limit expressions for $\chi$ predicted by existing theories were shown to be insufficient to explain the data.  A Weiss-like model based on the assumption that the rf response of the sample in the ferromagnetic state is dominated by the fluctuating, unpolarized fraction of the conduction band was shown to fit the data quite well.  It is hoped that this data will spur theoretical effort in understanding the dynamic properties of the ferromagnetic state of such systems.

\ack
We thank P. C. Canfield and G. Lapertot for supplying the samples, and J. Schmalian, P. C. Canfield, S. L. Bud'ko, G. Samolyuk, and V. Antropov for fruitful discussions. Work at the Ames Laboratory was supported by the Department of Energy, Office of Basic Energy Sciences under Contract No. DE-AC02-07CH11358. R.~P. acknowledges support from the Alfred P.~Sloan foundation. 


\section*{References}
\bibliography{magneticdb}

\providecommand{\newblock}{}
\begin{thebibliography}{10}
\expandafter\ifx\csname url\endcsname\relax
  \def\url#1{{\tt #1}}\fi
\expandafter\ifx\csname urlprefix\endcsname\relax\def\urlprefix{URL }\fi
\providecommand{\eprint}[2][]{\url{#2}}

\bibitem{myers-1999}
Myers K~D, Bud'ko S~L, Fisher I~R, Islam Z, Kleinke H, Lacerda A~H and Canfield
  P~C 1999 {\em J. Mag. Mag. Mat.\/} {\bf 205} 27--52

\bibitem{garrett-1981}
Garrett J~D, Greedan J~E and MacLean D~A 1981 {\em Mat. Res. Bull.\/} {\bf 16}
  145--148

\bibitem{ulrich-2002}
Ulrich C, Kahliullin G, Okamoto S, Reehuis M, Ivanov A, He H, Taguchi Y, Tokura
  Y and Keimer B 2002 {\em Phys. Rev. Lett.\/} {\bf 89} 167202

\bibitem{araki-2003}
Araki S, Metoki N, Galatanu A, Yamamoto E, Thamizhavel A and Onuki Y 2003 {\em
  Phys. Rev. B\/} {\bf 68} 024408

\bibitem{suzuki-1981}
Suzuki M and Ikeda H 1981 {\em J. Phys. Soc. Jap.\/} {\bf 59} 1133--1139

\bibitem{ho-1981}
Ho S~C, Maartense I and Williams G 1981 {\em J. Phys. F\/} {\bf 11} 699--710

\bibitem{drobac-1996}
Drobac D 1996 {\em J. Mag. Mag. Mat.\/} {\bf 159} 159--165

\bibitem{yelland-2005-2}
Yelland E~A, Hayden S~M, Yates S~J~C, Pfleiderer C, Uhlarz M, Vollmer R,
  v~L\"{o}hneysen H, Bernhoeft N~R, Smith R~P, Saxena S~S and Kimura N 2005
  {\em Phys. Rev. B\/} {\bf 72} 214523

\bibitem{vannette-2008}
Vannette M~D, Sefat A~S, Jia S, Law S~A, Lapertot G, Bud'ko S~L, Canfield P~C,
  Schmalian J and Prozorov R 2008 {\em J. Mag. Mag. Mat.\/} {\bf 320} 354

\bibitem{vannette-2008-2}
Vannette M~D, Bud'ko S~L, Canfield P~C and Prozorov R 2008 {\em J. Appl.
  Phys.\/} {\bf 103} 07D302

\bibitem{knapp-1971}
Knapp G~S, Fradin F~Y and Culbert H~V 1971 {\em J. App. Phys.\/} {\bf 42} 1341

\bibitem{vandegrift-1975}
VanDegrift C~T 1975 {\em Rev. Sci. Inst.\/} {\bf 48} 599--607

\bibitem{srikanth-1999}
Srikanth H, Wiggins J and Rees H 1999 {\em Rev. Sci. Inst.\/} {\bf 70} 3097

\bibitem{prozorov-2000}
Prozorov R, Giannetta R~W, Carrington A and Araujo-Moreira F~M 2000 {\em Phys.
  Rev. B\/} {\bf 62} 115

\bibitem{clover-1970}
Clover R~B and Wolf W~P 1970 {\em Rev. Sci. Inst.\/} {\bf 41} 617

\bibitem{prozorov-2006}
Prozorov R, Vannette M~D, Samolyuk G~D, Law S~A, Bud'ko S~L and Canfield P~C
  2006 {\em Phys. Rev. B\/} {\bf 75} 014413

\bibitem{dereotier-2006}
de~R\'eotier P~D, Lapertot G, Yaouanc A, Gubbens P~C~M, Sakarya S and Amato A
  2006 {\em Phys. Lett. A\/} {\bf 349} 513--515

\bibitem{stoner-1938}
Stoner E~C 1938 {\em Proc. Roy. Soc. Lond. A\/} {\bf 165} 372

\bibitem{moriya-1973}
Moriya T and Kawabata A 1973 {\em J. Phys. Soc. Jap.\/} {\bf 34} 639

\bibitem{mohn-2003}
Mohn P 2003 {\em Magnetism in the Solid State\/} (Springer)

\bibitem{wohlfarth-1968}
Wohlfarth E~P 1968 {\em J. App. Phys.\/} {\bf 39} 1061

\bibitem{saitoh-2004}
Saitoh E, Miyajima H, Yamaoka T and Tatara G 2004 {\em Nature\/} {\bf 432} 203

\bibitem{yelland-2005}
Yelland E~A, Yates S~J~C, Taylor O, Griffiths A, Hayden S~M and Carrington A
  2005 {\em Phys. Rev. B\/} {\bf 72} 184436

\end{thebibliography}

\end{document}